# Comparative Analysis of Impact of Cryptography Algorithms on Wireless Sensor Networks

Bilwasiva Basu Mallick and Ashutosh Bhatia

*Dept. of Computer Science and Information Systems*
*BITS Pilani, Pilani Campus*, INDIA

*Abstract*— Cryptography techniques are essential for a robust and stable security design of a system to mitigate risk of external attacks and thus improve its efficiency. Wireless Sensor Networks (WSNs) play a pivotal role in sensing, monitoring, processing and accumulating raw data to enhance the performance of the actuators, micro-controllers, embedded architectures, IoT devices and computing machines to which they are connected. With so much threat of potential adversaries, it is essential to scale up the security level of WSN without affecting its primary goal of seamless data collection and communication with relay devices. This paper intends to explore the past and ongoing research activities in this domain. An extensive study of these algorithms referred here, are studied and analysed. Based on these findings this paper will illustrate the best possible cryptography algorithms which will be most suited to implement the security aspect of the WSN and protect it from any threat and reduce its vulnerabilities. This study will pave the way for future research on this topic since it will provide a comprehensive and holistic view of the subject.

## I. INTRODUCTION

Wireless technology in the communications field along with power-saving design options have opened a new horizon of WSN which have expertly pertained to another parallel advancing domain of big data prediction, learning and analysis. With the aim for automation of industrial and production units, as well as the acceleration of MEMS and sensor technologies, the prospect of WSNs increasing rapidly.

The standardized and universally accepted definition of WSN is cited below from its Wikipedia page [1]. "A wireless sensor network consists of spatially distributed autonomous sensors to cooperatively monitor physical or environmental conditions, such as temperature, sound, vibration, pressure, motion or pollutants."

These small & inexpensive devices can effectively sense minute changes in the physical environment, monitor them for a specified time interval. Subsequently, it can communicate the findings to a common base station for further big data processing, computations and even future research options [36]. Such dealings of a vast volume of digital data is crucial to the functionality of a WSN. WSNs play a significant role in terms of complex real-time data collection, its dissemination and providing it for further processing [33].

Security of WSNs is a major concern due to its possible vulnerabilities to various types of cyber-physical attacks. The fact is WSNs are mostly deployed in harsh climatic regions for surveillance operations in the backdrop of challenging hostile circumstances. For example, it is used in enemy strongholds in war-zones, or even in a human-unreachable zone like icy mountaintops, dark and deep rainforests, unfathomable depths of ocean floor. As a result, several instances of malicious and threatening attacks will be possible for breaching the confidentiality, integrity and availability of information.

Not only this but also WSNs are integral parts in modern industries, medical healthcare services, meteorological departments, disaster management and relief operations, agricultural sector, civilian applications, smart home automation, environmental initiatives, traffic control and many other activities. Hence seamless data transmission by WSNs are required to ensure precision and accuracy of information. It is in such circumstances of open communication channels, the threat of information modification & compromisation, repudiation, replaying, masquerading, and other attacks happen.

The onus is to prevent such attacks for safety of devices and maintaining an overall integrity by managing a seamless protocol guaranteeing confidentiality, integrity, availability, authentication, authorization, two-way secrecies and non-repudiation.

## II. WSN SECURITY

This study aims to address the security issues of WSN, the degree of impact such potential attacks will yield and the various ways to counter them.

We can achieve the security of Wireless Sensor Networks by implementing Public Key Cryptography algorithms through efficient Key Distribution. But the deciding factor is the associated high complexity of such public key techniques which will render the very utility of the power efficient WSN completely useless.

A major step includes the identification of variational kinds of attacks on WSN in a bid to prevent any such future adversity. Security of WSNs is a major concern due to its possible vulnerabilities to various types of cyber-physical attacks.

The major constituents of the WSN device are sensor node, relay node, actor node, cluster head [basis of implementation of the LEACH algorithm routing protocol], entry and exit gateway and base station, which is situated in a remote location.

There is a myriad number of existing cryptographic algorithms being proposed and implemented. In this report, we make an exhaustive effort to explore all possible crypto algorithms and understand their individual effects on the performance of WSN.

The principal attributes of WSN which may cause the exposure of its vulnerabilities are random deployment, resource limitations, dynamic configuration, scalability, multi-hop inter-nodal communication. Maintenance of data privacy is a major concern in such WSN appliances where confidentiality is to be observed at every phase of perational lifetime.

The major quality attributes of a WSN which are to be enhanced for thwarting any breach in security are flexibility, resistance, self-organization, robustness, energy efficiency, self-configuration.

FIG.1 depicts the entire flow diagram of a WSN. Initially, the interconnected sensor nodes work in a dedicated manner amidst a typical environmental condition to monitor and collect the measurable raw data.

FIG.1 FLOW DIAGRAM OF WSN ARCHITECTURE

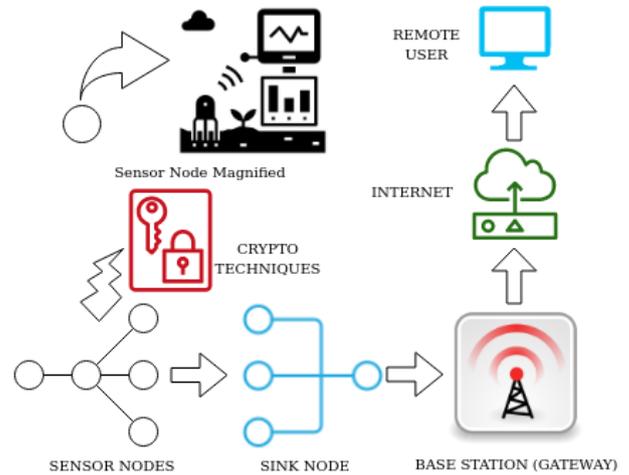

Following the collection, the nodes relay the information from source node to the sink node through relay nodes. Several routing algorithms are extensively studied to deduce the best optimized path for the proper operation of data travelling and data aggregation. The focus is made on maximizing network lifetime as well as minimizing the energy consumption. Since both these parameters are of conflicting interest, a trade-off should be made.

After data collection from the previous intermediate nodes, the sink node passes the data to the gateway node and the base station. The base station can be a satellite also which may collect navigation data in the Lower Earth Orbit (LEO) and send it to the appropriate ground station via downlink. If we consider the earth surface, an antenna can also act as a dedicated base station by taking care of uplink and downlink channels.

The functioning base station thus collects the data and uses the data forwarding operation to communicate it to the remote computing machine via internet cloud mechanism. A magnification of a singular sensor node has also been shown in FIG.1.

Due to such an extensive data transmission both within the WSN and outside it, the necessity of a secure channel arises. This particular example is taken from such existing literature. So, it is crucially important to deploy a cryptographic security protocol in the WSN operation stages, so that any unwanted externality is immediately thwarted.

Simulation of a typical WSN model with the interleaving of nodes and configuration of the entire

Network will be done in NS2, Omnet++ and Castalia. A comparative approach has been adopted to make an extensive study on the performance issues of WSN. When there are plenty of intelligent algorithmic approaches along with mathematical heuristics associated in each encryption methodology in cryptography, it needs to measure the applicability of such programming and design logic on the performance deciding parameters of WSN.

The attacks can be of two categories- the active and the passive attack. Passive attacks include snooping and traffic analysis. It also includes the Eavesdropping by passive listeners. Passive attacks are hard to detect. Even though they are not harmful, these can be decisive in the form of a hazardous information leaking. DoS and DDoS attack and attack on CIA [basic cryptosystem properties] form the active attack.

An illustration of a brief understanding of the various kinds of possible attacks on WSN is given in FIG.2.

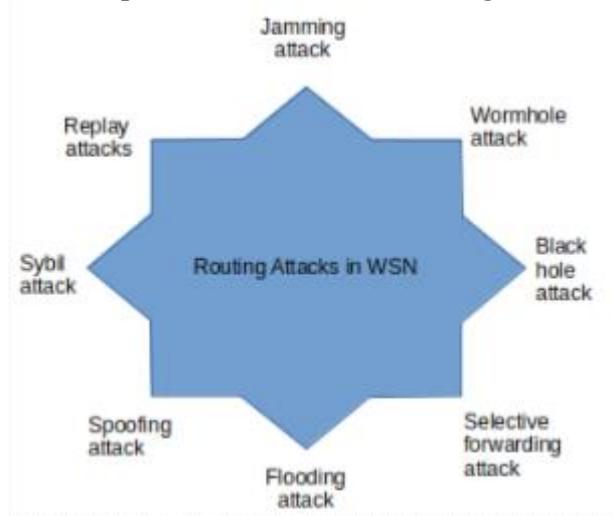

FIG.2 ROUTING ATTACKS IN WSN

There are several other possible attacks which include- Collision, Resource Exhaustion, Data Tampering, Sniffing, Hello, Flood, Replay, Impersonate. All of these attacks are further classified into the different groups based on the particular layers at which attacks are done. The classification has been tabulated in TABLE.1 and has further clarifications in Section IV.

### III. MOTIVATION

WSN plays a vital role in monitoring & recording physical & environmental data. In light of the unfortunate incident of Australian bushfire, a dedicated & efficient WSN would have led to an early detection & avoidance of the calamity. Being a resource-constrained device, scalability & power consumption is always a concern for WSN. Hence implementing additional security to WSN will increase the computational complexity and energy subsequently. So we need to present a smart and robust design of a properly functioning WSN which will use the hybrid of these two agendas of higher efficiency & enhanced security.

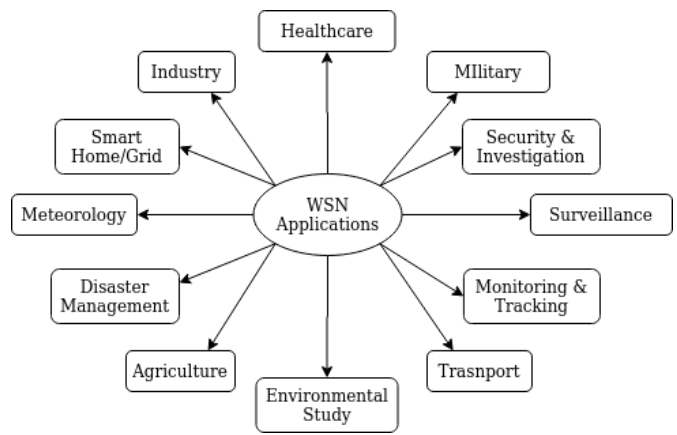

FIG.3 WSN APPLICATIONS

The vast domain of its potential implementations include the aspect of target tracking and investigation of extraterrestrial entities in other planets.

It is a major challenge in today's perspective. FIG.3 gives a vivid idea of the various fields in which WSNs find utilities. Such a versatile level of applications is what renders WSNs immensely beneficial.

### IV. OBJECTIVE

The project aims to observe the holistic effect on the main deciding parameters of a WSN- power consumption, energy efficiency & overall lifetime. To optimize the cost, we will use a wide range of cryptographic techniques- lightweight crypto, DES, AES and study the behaviour of the nodes & complete network. This experiment will provide an insight into the performance of the WSN under varying security architectures.

TABLE.1 CLASSIFICATION OF WSN ATTACKS

| PHYSICAL LAYER | DATA LINK LAYER | NETWORK LAYER [ROUTING PROTOCOL] | TRANSPORT LAYER |
|---|---|---|---|
| TAMPERING | UNFAIRNESS | SPOOFING | DE-SYNCHRONIZATION |
| JAMMING | COLLISION | REPLAYING | FLOODING |
| MESSAGE INJECTION | NODE EXHAUSTION | SINKHOLE | NODE SUBVERSION |
| PHYSICAL ATTACK OF NODE | | BLACKHOLE HELLO ATTACK | NODE COMPROMISE BLACKMAIL |
| | | WORMHOLE SLOWDOWN | |
| | | SYBIL ATTACK | |
| | | INFINITE LOOP | |
| | | SELECTIVE FORWARD [GREY HOLE ATTACK] | |
| | | IDENTITY REPLICATION ALTERING | |
| | | SLEEP DEPRIVATION TORTURE | |
| | | ACKNOWLEDGED SPOOFING | |

Hence we will be able to make a trade-off between improving the functionality of a WSN effectively & mitigate its risk to vulnerabilities. We will also check the impact of SHA-256 (hashing) & Elliptical curve cryptography

(RSA/Diffie-Hellman) techniques on our WSN virtual prototype to check their compatibility with the considerably low processing power & limited memory space of a WSN. Furthermore, we can come up with the best possible secured Routing Protocol in our WSN model based on our available design constraints.

## V. BACKGROUND STUDY

A comprehensive understanding of WSN & several Cryptographic techniques [conventional private key & public key; lightweight methods] has been done before proceeding to the core of the project.

Additionally, we have gone through the literature study of the undergoing research works in this area with all details of the state of the art technologies prevalent in this domain..

Table 1, details all the potential attacks in WSNs and puts them under the various subdivisions of the different layers of TCP-IP model. Most of the attacks fall under the Network Layer division since Routing plays a significant role in the architectural design and operation of the WSN. Memory constraint, limited power, energy exhaustion and lower computational capacity can make the WSN device prone to these detrimental attacks.

The proposed Cryptographic algorithms need to pertain to the requirements of a system. Since there are different kinds of adversaries happening in the various layers of the Network system, the established TCP/IP, the algorithms need to have better performance. Hence depending on the nature of the potential attacks, the developer needs to apply the correct algorithm.

A majority of the attacks are found to happen in the Network Layer, as it is vulnerable due to its working topology of routing and data forwarding- two integral functions of WSN. So a knowledge of such potential attacks can be helpful in adopting preemptive measures in getting the system to normalcy.

These attacks are very critical as far as the health of the device is concerned for its full operation capabilities.

For the application layer, there will be potential threats of Stimuli Attack and Packet Injection, which haven't been mentioned in the above table.

## VI. CRYPTOGRAPHIC METHODS

Several cryptographic methods are in vogue today. Some of them were developed in the very inception of the advent of the modern cryptography technology in the late 20th century.

Later, many of the techniques were inculcated recently with more modifications and optimizations of their previous counterparts.

A fundamental understanding of the cryptographic techniques are necessary to be acquainted with this particular problem statement of augmenting the security level of the WSN without compromising the performance metrics of its inherent parameters.

All the potential cryptographic techniques must comply with the quality attributes of a secured WSN- to confirm confidentiality, integrity, availability, data authentication, data freshness and non repudiation.

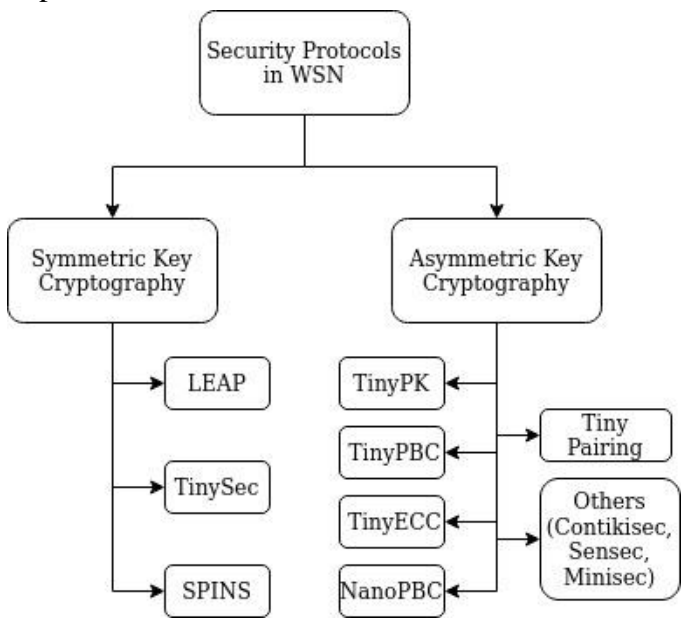

FIG.4. EXAMPLE OF WSN SECURITY PROTOCOLS

Symmetric key cryptography as well as hashing techniques will provide our cryptosystem with a preliminary protection for seamless data transmission by restoring and nurturing the qualities of integrity and confidentiality of the wireless communication medium.

This analysis will be further revisited in Section number XIV where the hybridisation of the algorithms will be discussed.

FIG.4. portrays the various security protocols of WSN which are in vogue presently.

The Public key cryptography is the emblem of data authentication and availability since it has the license of generating an efficient key distribution and management that guarantees scanning all the participating entities for any malicious activity. Hence any possible threat of an adversary, both inside the system or any external agent will be curbed down.

## VII. METHODOLOGY & OUTLINE

The complete flowchart of the working procedure is illustrated in FIG.5. This covers the design procedure to be followed in view of checking all the related crypto algorithms for a possible working potential in WSN. We use Castalia simulator platform for testing the proposed Cryptography algorithms on our WSN model.

Our project will be largely contributed by the utility of these algorithms as well as the conventional algorithms like AES, DES, Hash & MAC by feeding them in our WSN model & analyzing their behaviour.

Our preferred network simulator Castalia will help in modeling the virtual blueprint of a realistic WSN & studying its parameters under varying circumstances of different underlying cryptographic methods.

The project encompasses the domain of Cryptography & Network Security. It also covers a substantial portion of underlying Hardware architecture of WSN.

The working principles of WSN & its network functionalities will also be addressed.

Finally, the project will give a vivid understanding of the impact of the existing cryptography algorithms on a WSN.

## VIII. DATA AGGREGATION BASED CRYPTOGRAPHY IN WSN

One major challenge in the field of WSN is the maximization of the overall network lifetime of the device.

Data Aggregation can significantly reduce the energy dissipation of the system. It will help in achieving the desired energy efficiency.

Data aggregation can be defined as the method of collection and assimilation of data.

Data aggregation is one of the major steps in facilitating an energy-saving mechanism in WSN.

This technique forms the crux of the design principles.

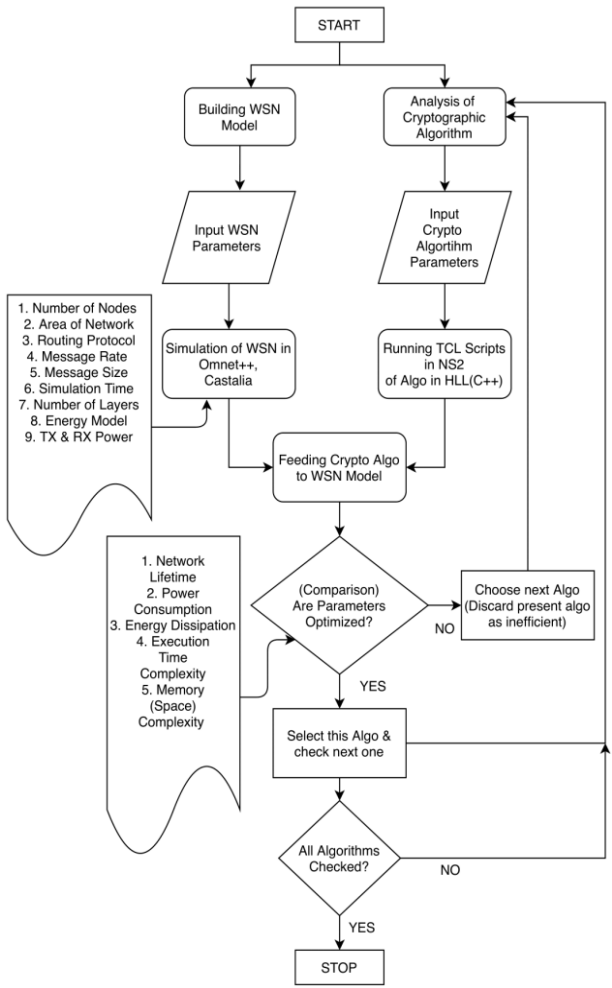

FIG.5. FLOWCHART OF ENTIRE WORKING DESIGN

In a nutshell, the three prime goals of this aggregation operation is to maximize the network lifetime of all nodes, energy efficiency of the device, and minimize the power dissipation of the system.

A secure energy-efficient data aggregation protocol called ESPDA (Energy-Efficient Secure Pattern based Data Aggregation) [35] can be used for Data Aggregation.

This method will keep a check on the redundancy of the system, as there will be no overlap in the transmitted packets.

Hence Data Aggregation is a key attribute in the efficiency of a WSN and even it can contribute to the enhancement of the security..

## IX. KEY DISTRIBUTION & MANAGEMENT

The various features which are associated with an efficient key distribution techniques in a typical WSN are pointed as follows:-
   a. scalability of the architecture
   b. optimization of energy consumption
   c. mitigation of power dissipation
   d. efficient memory utilization,
   e. increment of network lifetime
   f. reduction of computational cost
   g. maintenance of data capacity within the desired permissible limits
   h. renewal and resumption of the keyed activities
   i. smooth data exchange
   j. seamless connectivity between the interconnected nodes
   k. fortification of a strong defence against any possible attack.

There are some relevant papers in which contemporary Key distribution and management techniques are vividly illustrated in [3], [4], [5].

Key management techniques are efficient ways of organizing cryptographic materials and information in a cryptosystem. Key management technique in a whole includes the generation of keys, exchange of keys among sensor networks, storage of keys within sensor networks, uses and replacements of keys. It also includes cryptographic protocol design, key servers, interactive modules, user procedures and other relevant protocols in vogue. Such a task is a critical one, given the associated complexity of the whole procedure.

The authors in [30] have solely used Elliptic Curve Cryptography [ECC] without need of any digital signature. Such a certificate less ASK cryptography technique has been used in order to speed-up the process of encryption and decryption. This will reduce any overhead latency without much compromise of security aspect. If we compare and analyze the performance of the symmetric and asymmetric key cryptography techniques used in the different papers, then we conclude that the former significantly reduces the computation time of the cryptosystem. And there remains a loophole in protection against cryptanalytic attack, while the latter ensures unbreachable resistance to any kind of attack.

## X. LIGHTWEIGHT CRYPTOGRAPHY

The main crux of the word 'lightweight' cryptography reflects the adaptation of a simplified version of the standard techniques of cryptography by compressing the input sizes and other parameters in order to compensate for the power consumption of a WSN model. Furthermore, the hardware implementation of the 'lightweight' cryptography algorithms will be easier and it will be relatively easy to comply with the hardware architecture of WSN.
Lightweight cryptography is used extensively for devices with a critical constraint of using very low computing power. A tradeoff between light-weighted resource consumption and device security is the key factor here.
Researchers have come up with novel ideas to efficiently apply features of specific stream ciphers, block ciphers, hash functions and other encryption-decryption techniques to model such a robust design. In FIG.6 a design trade-off has been illustrated. Designers need to do manipulation of the security level and size & cost in order to scale-up the performance of the system. Hence the architecture should be balanced between serial or more optimized parallel versions.
The number of cryptographic rounds should vary between 8 and 48. And finally, the key size should be between 80 bits and 256 bits to nullify chances of any attack.

Such optimized lightweight techniques will render a device faster and will effectively reduce the overall lifetime of a WSN. Hence, the lesser time a WSN functions, the lower will be the power consumption. Lightweight devices have significantly lesser production cost and cover less space owing to its flexibility to be adjusted on the device surface.

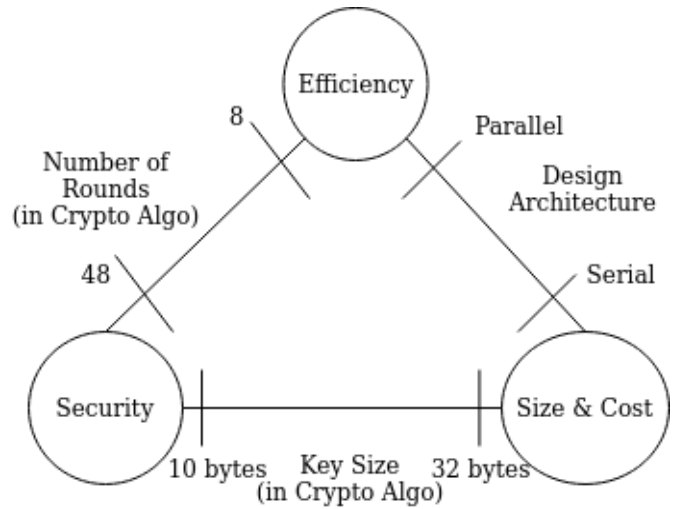

FIG.6. PARAMETER TRADEOFF OF LIGHTWEIGHT CRYPTO ALGORITHMS

In the following four sections, the different lightweight techniques in Block Cipher, Stream Cipher, Hashing methods and Hybridization Algorithms are extensively discussed in detail.

## XI. LIGHTWEIGHT BLOCK CIPHER TECHNIQUES

Block Ciphers are one of the fundamental methods used for a very simplified version of cryptography.
In this section, several contemporary lightweight versions of the existing Block Cipher methods are discussed and analyzed for their potential applicability in WSN devices as well as in M3WSN(Mobile Multimedia Wireless Sensor Network).

*A. AES*
The AES takes an input of 128 bits block length and a key size of 128 bits and operates 10 rounds on a typical substitution-permutation network [SPN]. AES can work with 128 bits, 192 bits or 256 bits key sizes. AES alone is not at all enough in providing security level to WSN.

*B. DESL and DESXL*
The dual algorithm methodologies proposed in [18] are modifications of the standard DES symmetric key algorithm. Unlike their predecessor DSA, these algorithms are easy to be implemented in Hardware, since they don't compute the initial and final permutations; and take only 1 S-Box. DESX is fed with an input of block length 64 bits and key size of 184 bits and it has 16 rounds of iteration on a fundamental Feistel structure.

Concerning DES, this algorithm is very much secure and has power saving properties. Since it is size-optimized in nature, it is prefered widely to applications that require stream ciphers.

*C. PRESENT*

The PRESENT algorithm [20] takes an input of 64 bits and a key size of 128 bits and produces the desired output after 32 rounds of iteration. The first 32 rounds consist of a linear permutation function, a non-linear substitution function and a XOR operation.

This algorithm is well protected against threats of slide attacks, associated keys. To be precise, the PRESENT algorithm is mostly used to cover up the limitations of AES. Hence PRESENT is a more optimized version of AES in light of working in a highly resource-constrained environment.

*D. CLEFIA*

This algorithm, proposed by SONY [6] is designed as a robust and secured version of the AES method. It easily aligns with AES since it works on an input of a block size of 128 bits and a key of length 128/192/256 bits is fed into it.

Experiments have shown that CLEFIA generates the maximum efficiency and the most optimum throughput for a key size of 128 bits and a fixed number of 18 cycles of rounds on the Feistel block structure. CLEFIA has a fast execution speed and is resistant towards differential and linear cryptanalysis threats.

CLEFIA generates better performance than the PRESENT algorithm in terms of throughput. However, PRESENT outperforms CLEFIA in overall security and memory utilization.

*E. KATAN*

This algorithm, proposed by [26] uses block ciphers which are applied on streams to generate the encrypted output. It takes an input of a block size of variant lengths of 32, 48, 64 and a key size of 80 bits and operates on a stream structure iterating over 256 rounds for encryption. The streaming concept arises from the loading of the encrypted processed blocks into a pair of shift registers which further acts as dedicated feedback registers to augment the utility of nonlinear functions generated during the 256 rounds.

*F. HIGHT*

This algorithm, proposed by [19] operates on an input of block length of 64 bits and a key size of 128 bits. It performs an iteration of a total 32 rounds on a Feistel block structure. HIGHT has an easy implementation and supports the embedded architecture of the WSN.

The HIGHT algorithm facilitates ultra-lightweight mechanisms and hence its performance is quite detrimental with respect to the classical block ciphers.

*G. KLEIN*

KLEIN algorithm proposed by [27] is fed with an input of a block size of 64 bits and an optional key size of 64/ 80 /96 bits. This method operated on a Substitution-Permutation network and has 12/ 16/ 20 rounds of iteration.

Differential cryptanalysis attack is nullified in this process but there will be chances of integral attack in the system.

*H. TWINE*

TWINE algorithm proposed earlier [21] is operated on a 64-bit block cipher and a variable key size of 80-bits or 128-bits on a Feistel structure of type-2. Earlier, this design lacked a solid diffusion property. Later it was further modified by [7] by replacement of the Feistel structure.

The enhanced version when operated with a complete iteration scheme, will thwart any possibility of the Meet in the Middle attack and will also produce faster encryption since the key structure generates unique random numbers to stall any slide attack.

There are several other available algorithms. All of the algorithms have been thoroughly studied by the author and represented in a tabular manner to increase the comparative analysis amongst all the block cipher models.

These algorithms are computationally lightweight and will perform decently for the security application procedures in the WSN.

TABLE.2 TABULAR REPRESENTATION OF BLOCK CIPHERS

| NAME OF CIPHER | STRUCTURE | BLOCK LENGTH | KEY SIZE | NUMBER OF ROUNDS |
|---|---|---|---|---|
| LED | SPN | 64 | 128 | 32/48 |
| AES-128 | SPN | 128 | 128 | 10 |
| PRESENT | SPN | 64 | 80, 128 | 31 |
| NOEKEON | SPN | 128 | 128 | 16 |
| MCRRYPTON | SPN | 64 | 96, 128 | 12 |
| PRINCE | SPN | 64 | 128 | 12 |
| PRINT | SPN | 48,96 | 80,160 | 96 |
| KLEIN | SPN | 64 | 80, 96 | 12/16/20 |
| BSPN | SPN | 64 | 64 | 8 |
| CLEFIA | FESITEL | 128 | 128 | 18 |
| DESXL | FESITEL | 64 | 184 | 16 |
| SKIPJACK | FESITEL | 64 | 80 | 32 |
| LBLOCK | FESITEL | 64 | 80 | 32 |
| HIGHT | FESITEL | 64 | 128 | 32 |
| PICCOLO | FESITEL | 64 | 80, 128 | 25/ 31 |
| SEA | FESITEL | 96 | 96 | VARIABLE |
| TWINE | FESITEL | 64 | 80/128 | 36 |
| TEA & XTEA | FESITEL | 64 | 128 | 64 |
| KATAN & KTANTAN | FESITEL | 32 | 64/80 | 254 |
| MIBS | FESITEL | 64 | 80/128 | 32 |
| IDEA | LAI-MASSEY | 64 | 128 | 8.5 |

Based on further research studies and other articles related to the industrial functionality of WSN, it has been evident that PRESENT, CLEFIA and HIGHT are used in the operations of WSN, as far as applicability of block ciphers is concerned.

*I. PRINT*

The authors in [31] have come up with a symmetric block cipher Print, which is used in printing Integrated Circuits (IC). It is fed with an input of block length of 48 or 96 bit and a key size of 80 or 160 bits on a SPN structure. It performs XOR operation with the round key and iterates over 96 rounds. This technique is resistant to algebraic attacks, related key attacks and higher order differential attacks

TABLE.2 gives a vivid description of all the algorithms with their parameters to provide a concrete platform for trade-off and analysis. SPN refers to the Substitution Permutation Network.

All the block cipher techniques are covered here along with a tabular data of their key size, block length and number of rounds. We have discussed 8 specific techniques out of the 20 examples we have tabulated.

It has been experimentally proven that the overall energy efficiency of WSN is better for the block ciphers than the stream ciphers.

Overall, the various parameters of Block Cipher which play a pivotal role in determining the energy consumption of a WSN are as follows:-
  A. Selection of Initialization Vector
  B. Number of rounds of iteration
  C. Key Size & Block Length
  D. Key Selection & Setup

## XII. STREAM CIPHERS

Since Stream ciphers are rarely used in WSN operations hence, only some references are cited in [2] and [8] which have worked on chaotic stream ciphers.

In [2] the authors have designed a secure cryptosystem that will guarantee the feasibility of data encryption between all respective nodes of WSN. It also significantly takes care of additional power consumption and device cost and provides defense against MITM and replay attacks.

In [8] the authors have designed a stream cipher eLoBa which reduces transmission overhead and performs well against algebraic attacks. This technique can rule out any chance of data alteration and tampering by any unwanted malicious external agent.

TABLE.3 TABULAR REPRESENTATION OF STREAM CIPHERS

| Stream Cipher | Key-stream size (byte) | Number of cycles |
|---|---|---|
| Sosemanuk | 80 | 8559 |
| Trivium | 80 | 80 |
| Salsa | 64 | 17812 |
| Grain | 80 | 80 |
| RC4 | 1 | 31 |

However, it has been found that stream ciphers do not really pertain to the structural and functional necessity of WSN.

Hence both of these proposed methodologies work well in case of fringe applications of multimedia data streaming or other broadly classified wireless technologies. WSNs are much more resource-constrained.

Tahir et. al. also proposed a lightweight authenticated encryption mechanism combined with hashing methods and based on Rabbit stream cipher for WSN application.

For further clarification, a tabular representation of the stream ciphers has been cited in TABLE.3.

As it is evident that RC4 will use the least number of CPU cycles per bytes of stream data, it will be able to reduce latency.

Theoretically also, it is proven that RC4 has the best possible execution time amongst all other symmetric and asymmetric ciphers and only an ECC technique can outperform it.

### XIII. HASHING

MD5, SHA1, RIPEMD are some famous cryptographic hashing based algorithms which uphold the quality of message signature techniques. However hash functions suffer from severe overhead problems and also the encryption aspect is not much emphasized in hashing based technique, as reflected in Message Authentication Code [MAC] and other random key based hashing principles. The hashing methodology mainly harnesses the attribute of integrity within the cryptosystem.

In [9] the authors have devised a lightweight preimage and collision-resistant one-way hashing function without any key structure. This design gives a lower overhead cost for the MD5 and SHA1 techniques. However the design does not have an experimentally proven application in WSN since message authentication options and real-time functionality features are yet to be introduced. Since MAC involves the accessibility of a key based structure, hence the system shall not be ideal for WSN applications.

The author in [25] has also designed the same version of the algorithm as in [9] with a one-way hashing technique that gave better overhead.

In [23] and [24] the authors have used MAC based cryptosystem generation which guarantees data authentication & message integrity and prevents replaying attack respectively.

HMAC (Hash based Message Authentication Code) is another novel technique where the key sze of the key will be the deciding factor for finding how much secure the cryptosystem is.

It has also been used in [22], where a pseudo-random number generator function and the key is used. Here, data authentication is supported. Privacy of data in one node is preserved since neighbouring nodes and external eavesdroppers can't do any trespassing or snooping.

### XIV. HYBRID CRYPTOGRAPHIC ALGORITHMS

Since WSN requires a robust mode of security protocol, a blending of symmetric and asymmetric key techniques is performed to capture the essence of optimum security assurance of public key cryptography and the non-complex easy-to-design attributes of a private key cryptography technique.

[13] have come up with a novel idea of a three-pronged mega security proposal which addresses the desirabilities of confidentiality [Elliptical Curve Cryptography (ECC) and Advanced Encryption Standard (AES)], integrity [using Message Digest-5

(MD5)] and authentication [XOR-DUAL RSA algorithm].

As far as cipher text size, WSN energy consumption, encryption & decryption computation time are concerned this hybrid algorithm outperforms the existing ones. Hence this methodology efficiently reduces the individual overhead of nodes and collectively reduces the energy dissipation.

Although Symmetric key Cryptography techniques enhance security and are cost-effective in nature in terms of time and space, they have the limitations of constraints associated with sharing the private key between the parties.

Asymmetric key Cryptography techniques provide encryption schemes by efficient key distribution but are computationally expensive on the other hand. Hence their hybridisation is the best possible option to boost up the advantage of each one of them. Since there is utility of two keys cumulatively key space analysis confirms full security against malicious brute force attacks. So, such techniques find immense application to make image encryption based cryptosystems literally invincible to any kinds of adversaries.

In the first phase, confidentiality & information security are embedded by a combination of ECC & AES, followed by availability and user authentication implemented by XOR-DUAL-RSA in the second phase. And overall hashing based MD5 provides the advantage of message signature technique to guarantee integrity of the designed system.

[16] designed a hybrid algorithm based on sequential encryption of AES and ECC techniques. It provides preliminary security options for a spatially distributed ad-hoc wireless communication system of autonomous sensor nodes. However, it is also likely to spend a higher execution time for its completion.

[17] devised a combination of DES and RSA to address the block cipher encryption and the key distribution procedure. However the security was fragile and easily vulnerable to cryptanalysis with less key sizes and overall it is slower compared to a singular symmetric key implementation.

The authors in [28] have come up with a hybridisation of ECC and RC4 algorithms. It works by harnessing the Elliptic Curve Cryptography (ECC), an asymmetric key cryptography, for efficient key distribution to address the memory optimization concern. It further uses the RC4, a stream cipher, since it speeds up the encryption and decryption process. Although it draws lesser power and improves throughput, the level of security remains a concern.

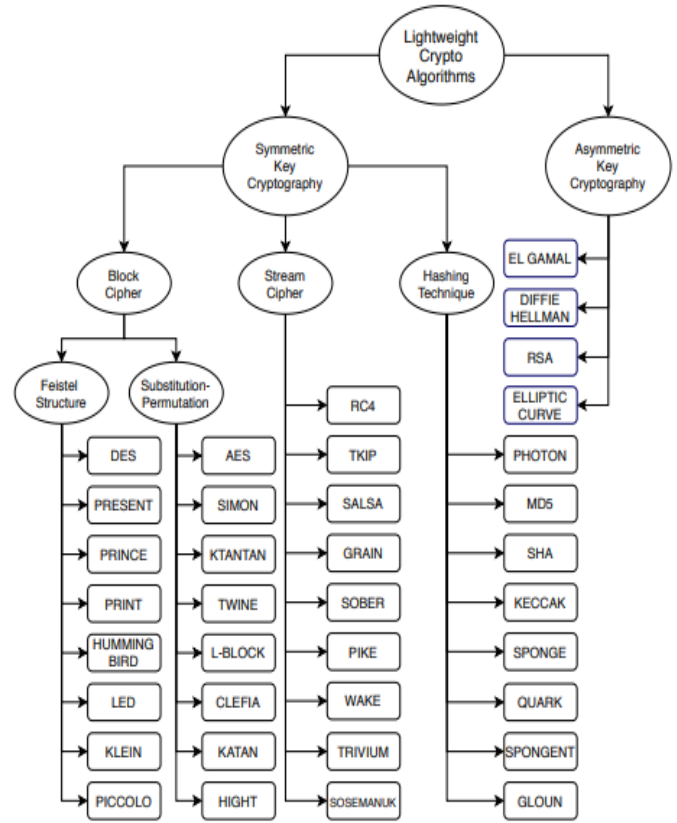

FIG.7. CLASSIFICATION OF LIGHTWEIGHT CRYPTO ALGORITHMS

In [29] the authors have proposed a three-stage hybridisation technique to minimize the execution time. They have applied Advanced Encryption Standard (AES) in the initial phase, followed by Data Encryption Standard (DES) in the next round. Once these two successive rounds ensure security, a modified version of Rivest–Shamir–Adleman (m-RSA) is used in the final round. The last round provides a simplified key distribution option. Due to their distributed execution across the three stages, the parallelization is established. It gives better results as compared to each singular algorithm when executed alone.

## XV. PUBLIC KEY CRYPTOGRAPHY

The asymmetric key techniques are also an integral part of the different techniques used in WSN models. Compared to the private key technologies these algorithms have a better model for secrecy and authentication of information. Key distribution schemes and provision of digital signature methods are the hallmarks of these algorithms.

Tiny-PK is a lightweight version of a RSA cryptosystem which requires digital certificate authentication based on the Diffie-Hellman key agreement protocol.

Owing to its relatively smaller size, Elliptic Curve Cryptography (ECC) is often used for WSN applications. The key distribution scheme of this particular method is computationally lightweight with a higher efficiency.

[32] provides a lucid analysis of a trade-off between RSA and ECC methods of Public Key Cryptography. ECC edges over its RSA [Rivest, Shamir and Adleman] counterpart in terms of lesser memory complexity. The time taken for encryption and decryption steps for a fixed key size is lesser for ECC. Even the more optimized versions of RSA- the one with Chinese Remainder Theorem implementation, is also outperformed by ECC. A key size of 60/224/256-bits for ECC has the potential of the same security level as that of a 1024/2048/3072-bit sized RSA key. The much celebrated Multi Prime RSA, another version of the algorithm, has poor time and operational efficiencies that that of ECC. The authors of [34] has also described the advantage of ECC over RSA or other symmetric key techniques in vogue. The idea is regarding improvement of models using modified coordinate techniques, elastic windows to thwart Special Power Analysis [SPA] attacks. TINY-ECC technique is a lightweight mechanism which provides a digital signature scheme, a key agreement protocol and a public key encryption operation. However this also has a drawback of lower scalability.

Table 4 depicts the problems of several public key crypto algorithms. All of these shortcomings with respect to potential application in WSNs paves the way for more utility of ECC.

TABLE.4 TABULAR REPRESENTATION OF PUBLIC KEY CRYPTO TECHNIQUES

| Name of Public Key Cryptography Algorithm | Problem in WSN Application |
|---|---|
| Chor-Rivest Knapsack | Subset Sum |
| Digital Signature Algorithm [DSA] | High Time Complexity |
| RSA | Memory Complexity |
| Mc Eliece | Linear Code Decoding |
| El Gamal | Diffie Hellman |
| Blum Goldwasser | Space Complexity |
| Rabin | Integer Factorization |

This will guarantee a higher freshness, node authenticity, data integrity and confidentiality in the embedded architecture design.

However with the further evolution of Quantum cryptography this public key crypto algorithms are under more challenge due to latest research implementations in quantum computing – namely Shor's algorithm.

## XVI. BEST PRACTICES OF CRYPTOGRAPHY METHODS

There are several more crypto techniques which haven't been discussed here. This is an evolving field and so many new introductions of novel methods are possible here. So, upto this section we have provided a comprehensive review of the merits and demerits of the existing algorithms.

Here is a summary of the utility of these algorithms as far as the application of WSN is concerned.

FIG.10 depicts a pie chart which signifies the percentage of utility of these algorithms in the application of today' security architectures in WSN.

## XVII. CONCLUSIONS

This paper has addressed the potential algorithmic techniques which can be adopted to upgrade the security level of the Wireless sensor networks.

The results converge to a concrete conclusion that the private key cryptographic techniques as well as the hashing methods are widely used in the WSNs.

In this paper we have elucidated on a state of the art in the domain of Cryptography and Network Security. It encompasses the numerous algorithms meant to be implemented on a WSN in order to reduce the chances of any kind of attack. We have covered the ins and outs of the various attacks that pose inimical threats and can potentially damage the objectivity of the sole purpose of deployment of the WSN.

Throughout all the previous sections, we have methodically described a synthesis of all possible cryptography systems and the underlying algorithms which will consolidate the utmost safety of the WSN. Hypothetically speaking, it may be possible to build a cryptosystem that has all the desired qualities of supporting in the WSN configuration. But as of now, for the time being, there seems not to be a complete and dynamic system which can single-handedly account for cent percent security of the WSN cryptosystem.

The project will aim to achieve a statistical comparison of the efficiencies of our WSN under the different crypto algorithms. This will result in a crystal clear & comprehensive idea of additional aspects of WSN- key management & distribution, data provenance under the optimal algorithm. Apart from an engaging learning opportunity in partial fulfillment of NETWORK SECURITY [CS G513] course, the project will provide substantial knowledge on the intriguing domain of WSN- its range of parameters, behaviour under varying conditions. Further studies & simulations will lead to a concrete idea of the working methodologies of the crypto techniques in WSN.

In the near future, the plan is to write TCL scripts in NS2 and run these existing algorithms on a simulated WSN for getting a more concrete proof of our idea. As a precursor, we have explored several papers on simulations which have carried out these experiments and concluded their result. We will soon design our only simulation and run the related scripts to check the algorithms. We have already designed the WSN parameters in pen and paper. We have chalked out the methodologies to synthesize our observations and come up with a concrete inference.

The major outcome and the key takeaway of this research is the supremacy of Private Key cryptography algorithms over their asymmetric public-key counterparts. Along with the different lightweight Hashing techniques it is thoroughly established that these techniques are robust and have enough level of security to address any challenge. It is very much appropriate for utility in the domain of WSNs. Also amongst the public key techniques, ECC has the best efficiency in terms of lesser space and time complexity as well as better security performance, as compared to the other techniques with the same

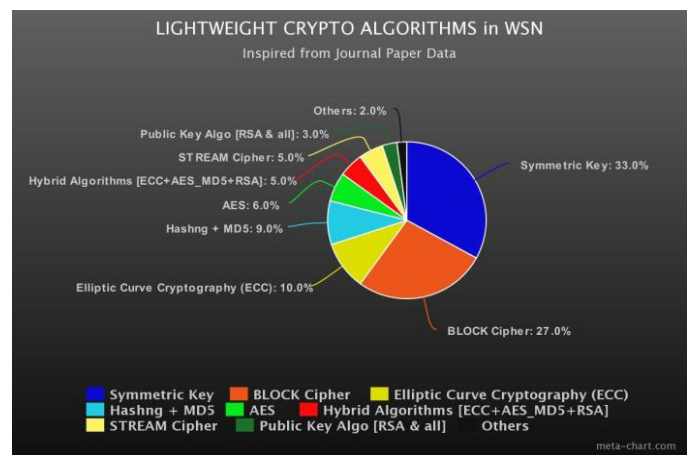

FIG8. PIE-CHART REPRESENTATION OF THE UTILITY OF LIGHTWEIGHT CRYPTO ALGORITHMS IN WSNs

Hence the key takeaway is the utility of the symmetric key cryptography technique and a hybridisation of Hashing technique to optimize the authentication attribute of the system. For mission critical cases, only the former will be enough.